# A simplified procedure to numerically evaluate triggering of static liquefaction in upstream-raised tailings storage facilities

Mauro Giuliano Sottile
Universidad de Buenos Aires and SRK Consulting, Buenos Aires, Argentina, msottile@srk.com.ar

Ignacio Adolfo Cueto
SRK Consulting, Buenos Aires, Argentina, icueto@srk.com.ar

Alejo Oscar Sfriso
Universidad de Buenos Aires and SRK Consulting, Buenos Aires, Argentina, asfriso@srk.com.ar

ABSTRACT: The interest of the mining industry on the assessment of tailings static liquefaction has exacerbated after recent failures of upstream-raised tailings storage facilities (TSF). Standard practices to evaluate global stability of TSFs entail the use of limit equilibrium analyses considering peak and residual undrained shear strengths; thus, neglecting the work input required to drive the softening process that leads to progressive failure of susceptible tailings. This paper presents a simplified procedure to evaluate the static liquefaction triggering of upstream-raised TSFs by means of finite element models employing the well-known Hardening Soil model with small-strain stiffness (HSS). A calibration methodology is proposed to overcome the model limitation of not being implemented in a critical state framework, focusing on the stiffness parameters that control the rate of shear-induced plastic volumetric strains. A real TSF is modelled in Plaxis 2D to evaluate its vulnerability to liquefy due to an undrained lateral spreading at the foundation. Results show that minor movements near the toe induce the material into a strain-softening regime that leads to a progressive failure towards the structure crest.

KEYWORDS: Static Liquefaction, Triggering, Tailing Storage Facilities, Plaxis 2D, HSS model.

RESUMO: O interesse da indústria de mineração na avaliação de liquefação estática em rejeitos intensificou-se após as recentes falhas em barragens de rejeitos alteadas a montante (TSF). Os procedimentos padrão para avaliação da estabilidade de TSFs envolvem análises de equilíbrio limite considerando resistências não drenadas de pico e residual; negligenciando, assim, o trabalho necessário para iniciar o processo de amolecimento que provoca falhas progressivas em rejeitos. Este artigo apresenta um método simplificado para avaliação dos mecanismos desencadeadores de liquefação estática em barragens alteadas a montante através de modelagem numérica e do modelo Hardening Soil com pequenas deformações (HSS). É proposta uma metodologia de calibração para superar a limitação que o modelo apresenta por não estar implementado segundo uma teoria dos estados críticos, com foco nos parâmetros de rigidez que controlam a taxa de deformação volumétrica plástica induzida por cisalhamento. Um TSF real é modelado no Plaxis 2D para avaliar a sua vulnerabilidade à liquefação estática gerada pelo deslizamento não drenado da fundação. Os resultados mostram que pequenos movimentos ao nível do pé induzem um regime de amolecimento no material que provoca uma falha progressiva em direção à crista da estrutura.

PALAVRAS-CHAVE: Liquefação Estática, Mecanismos Desencadeadores, Barragens de Rejeitos, Plaxis 2D, modelo HSS.

## 1   Introduction

Tailings are mining waste products obtained from rock crushing which are normally disposed as a slurry into storage facilities (TSFs); thus, the material generally has a loose in-situ structure due to the lack of post-deposition compaction and the electrical interaction among finer particles. The combination of contractive states with high degrees of saturation -mainly during the operation of the facility- induce the material into a susceptible condition against static liquefaction, for which a post-peak strain-softening response is expected for undrained shearing; the brittleness of the behavior can be increased if particle arrangements are locked by early diagenesis cementation after placement (Santamarina et al, 2019). The upstream construction method for TSFs has been widely used in the industry, as it is highly convenient from a financial perspective; however, it has proven to be more risky than other methods, as the global stability relies on the susceptible (i.e. loose and saturated) tailings strength.

Recent failures -such as Merriespruit, Mount Polley, Samarco and Brumadinho- have exacerbated the industry interest in the assessment of static liquefaction of TSFs. Standard practices to evaluate global stability entail the use of limit equilibrium analyses considering peak and residual undrained shear strength; thus, neglecting the work input required to drive the softening process that leads to progressive failure of susceptible tailings. Deformation modelling with an appropriate constitutive model formulated in effective stresses can account for many of the complexities of the tailings behavior that cannot be captured using limit equilibrium techniques, including the post-peak strain-softening response for undrained loading. This paper presents a simplified procedure to evaluate static liquefaction triggering of upstream-raised TSFs by means of finite element techniques and the well-known Hardening Soil model with small-strain stiffness (HSS). A calibration methodology is proposed to overcome the model limitation of not being implemented in a critical state framework. An application to a real TSF is presented to evaluate its vulnerability to liquefy for a lateral spreading mechanism at the foundation.

## 2   Calibration methodology for HSS constitutive model

### 2.1   Background

It must be acknowledged that there is still a gap between constitutive models used in the industry and those developed for academical purposes, which have clear capabilities to reproduce tailings complex behavior, but generally require high quality laboratory data and are seldomly implemented in commercial softwares. To illustrate: NorSand model (Jefferies, 1993) is formulated in a critical state framework and entails a simple calibration process, it has excellent capabilities to reproduce CIUD|CIUC tests on sand/silt specimens, but it falls short on reproducing direct simple shear modes which is crucial for slope stability modelling (Castonguay, 2018). On the other hand, a more sophisticated model, such as the MIT-S1 model (Pestana & Whittle, 1999), has great capabilities on capturing anisotropic stress-strain-strength behavior of clay/silts/sands and is an optimum tool to assess static liquefaction, but it unfortunately requires high quality laboratory testing that is generally not available in routine assessments. In an attempt to reasonable capture susceptible tailings undrained behavior using limited testing data and a commercially available -and industry accepted- constitutive model, a calibration methodology for the HSS model is presented in this study; it must be highlighted that the purpose is not to replace the role of the existing models, but to provide a methodology for practical static liquefaction assessments using tools widely available in geotechnical engineering practice.

### 2.2   Model formulation

HSS is an effective stress isotropic hardening plasticity model that is the flagship of Plaxis. The model is not implemented in a critical state framework, as the Critical State Line (CSL) is not defined in terms of $p' - q - e$ because void ratio is not a state variable; thus, the stress-strain-strength response predicted by the model for undrained shearing does not depend on the material initial density, which is a main limitation for sands and silts. To account for this shortcoming, a calibration methodology is proposed focusing on the stiffness parameters that control the rate of shear-induced plastic volumetric strains, which has a direct impact on the pore pressure induced during undrained shear and so, on the peak and residual undrained shear strengths.

The simplest version of HSS employ Mohr-Coulomb failure criterion and inherits most of the Hardening Soil (HS) model features (Schanz et al, 1999), including a cap-type yield surface that accounts for plastic volumetric strain in proportional compression, and a shear hardening yield surface that evolves with plastic shear strain. The latter is a key component of the proposed calibration methodology, and its yield function expression is:

$$f_s \cong \frac{2-R_f}{E_{50}} \cdot \frac{q}{1-q/q_a} - \frac{2 \cdot q}{E_{ur}} - \bar{\gamma}^p \tag{1}$$

where $q$ is the deviatoric stress, $R_f$ is a material parameter, and $q_a$, $E_{50}$ and $E_{ur}$ are functions of material parameters and stress states. The expression shown in Eqn. (1) is formulated from experimental observations of drained triaxial monotonic tests, where: the first term is twice the total axial strain ($\epsilon_1$), and is related to the deviatoric stress by a hyperbolic formulation; the second term is twice the elastic axial strain ($\epsilon_1^e$); and $\bar{\gamma}^p$ is a hardening parameter that accounts for accumulated plastic shear strain. The reader is referred to Schanz et. al. (1999) and Benz (2006) for details of the HSS model and a full definition of all variables employed.

The model uses a non-associative shear flow rule that linearly relates the rate of shear-induced plastic volumetric strain ($\dot{\epsilon}_v^{ps}$) and the rate of plastic shear strain ($\dot{\gamma}^p$), given by the mobilized dilatancy angle ($\psi_m$). Note that $\bar{\gamma}^p$ and $\dot{\gamma}^p$ have different notations to highlight that they are two different things: one is a state variable and the other one is a measure of the plastic shear strain increment. The relationship between $\dot{\gamma}^p$ and $\dot{\epsilon}_v^{ps}$ is:

$$\dot{\epsilon}_v^{ps} = \sin\psi_m \cdot \dot{\gamma}^p \tag{2}$$

In turn, $\psi_m$ depends on the mobilized friction angle ($\phi_m$), or equivalently, on the stress ratio ($\eta = q/p'$). For contractive materials undergoing undrained loading, the range of interest to reproduce plastic strain evolution is $0 < \phi_m < \phi_{cv}$, where $\phi_{cv}$ is the constant volume friction angle, an internal parameter of the model. Dilatancy in HSS model is computed using a modified version of Li & Dafalias (2000) formulation, presented in Eqn. (3), where: $q/q_a = [(1 - \sin\phi_{cv})(\sin\phi_m)]/[(\sin\phi_{cv})(1 - \sin\phi_m)]$; $M_c = 6 \cdot \sin\phi_{cv}/(3 - \sin\phi_{cv})$; $M_d = 6 \cdot \sin\phi_m/(3 - \sin\phi_m)$. In order to avoid unrealistic volume changes at low stress ratios, Plaxis implementation of HSS defines a cut-off value for $\sin\psi_m$ by limiting $\sin\phi_m$ to a minimum value of $\sin[\phi_m^{min}] = \sin\phi_{cv}/(2 - \sin\phi_{cv})$ when computing $M_d$ in the range $0 < \phi_m < \phi_m^{min}$. The dilatancy formulation of HSS is one of the main differences with the HS model.

$$\sin\psi_m = \frac{1}{10}\left\{-M_c \exp\left[\frac{1}{15}\ln\left(\frac{M_c}{M_d}\frac{q}{q_a}\right)\right] + M_d\right\} \tag{3}$$

### 2.3 Calibration for CIUC tests

The shear hardening yield function presented in Eqn. (1) depicts that stiffness functions $E_{50}$ and $E_{ur}$ play a key role on the evolution of the shear hardening parameter $\bar{\gamma}^p$. The rate of shear-induced plastic volumetric strain and, consequently, the shear-induced pore pressure and undrained shear strengths are controlled by these two functions. Then, the behavior of a very loose material state can be reproduced with low values for $E_{50}$ and high values for $E_{ur}$, while the response for denser states can be captured by just increasing $E_{50}$. It must be noted that these two are just hardening functions, not fundamental material properties. Both are stress dependent through power laws, and therefore evolve during shearing; they are calibrated at a reference pressure ($p_{ref}$) by two material parameters, namely $E_{50}^{ref}$ and $E_{ur}^{ref}$. The calibration of HSS is initially done using CIUC test data; the aim is to find a set of parameters that can reasonable reproduce the undrained response of the material for different pre-shearing states by adjusting $E_{50}^{ref}$.

In this paper, data from three CIUC tests performed on reconstituted low-plasticity silt-like tailings samples is used. The samples are consolidated to $\sigma_3' = 400\ kPa$ with pre-shearing void ratios $e_0 = [0.61, 0.69, 0.71]$; with equivalent state parameters $\psi = [0.04, 0.12, 0.16]$ when computed using a semi-log CSL with $\Gamma = 1.1$ and $\lambda_{10} = 0.205$. The calibration strategy for the various parameters is: i) cohesion ($c'$) and friction angle ($\phi'$) are determined from the failure envelope; ii) the small-strain shear modulus ($G_0^{ref}$) is determined from geophysics testing; iii) unload-reload and oedometric reference stiffnesses ($E_{ur}^{ref}$ and $E_{oed}^{ref}$) are calibrated from CSR tests; iv) parameters $K_0^{NC}, R_f$, threshold shear strain ($\gamma_{0.7}$), stress exponent ($m$), and Poisson's ratio ($\nu_{ur}$) are adjusted to fit the CIUC data, while $E_{50}^{ref}$ is adjusted to capture the undrained response for each pre-shearing density. A summary of the calibrated parameters is shown in Table 1. Using these parameters, a comparison between the three CIUC data and elemental tests is shown in Figure 1.

Table 1. HSS calibrated parameters for tailings.

| Parameter | $c'$ | $\phi'$ | $G_0^{ref}$ | $\gamma_{0.7}$ | $p_{ref}$ | $R_f$ | $K_0^{NC}$ | $m$ | $\nu_{ur}$ | $E_{ur}^{ref}$ | $E_{oed}^{ref}$ | $E_{50}^{ref}$ |
|---|---|---|---|---|---|---|---|---|---|---|---|---|
| Units | kPa | ° | MPa | - | kPa | - | - | - | - | MPa | MPa | MPa |
| Value | 0 | 35 | 50 | 1E-4 | 100 | 0.875 | 0.6 | 0.55 | 0.25 | 60 | 9.0 | 3.0\|4.5\|9.0 |

A remarkable agreement in terms of effective stress paths, peak/residual undrained shear strengths, and residual shear-induced pore pressures is achieved for the three tests; however, the model falls short in capturing the strain-at-peak and the evolution of pore pressures with strain, which is a direct consequence if using low values of $E_{50}^{ref}$. The HSS shear hardening yield surface (SHYS) loci for constant values of accumulated plastic shear strains of 1, 2, 5, 10 and 20 % are plotted in green dotted lines; it is seen that the loosest sample accumulates high plastic shear strains from early test stages, while opposite occur for the denser sample. The HSS initial cap hardening yield surface (CHYS) is plotted in dotted blue lines; it is observed that the loosest sample calibration has the steepest cap configuration, which minimizes the cap influence during shearing. In addition, the HSS isotropic equivalent stress ($p_{eq}$) and preconsolidation pressure ($p_p$) evolution with strains are plotted in grey lines; it is observed that the loosest samples do not experience any volumetric hardening, as $p_p$ remains constant, while the densest sample has some cap-hardening before reaching failure; this is key, as demonstrates that loose samples undrained behavior is totally controlled by the shear hardening yield surface.

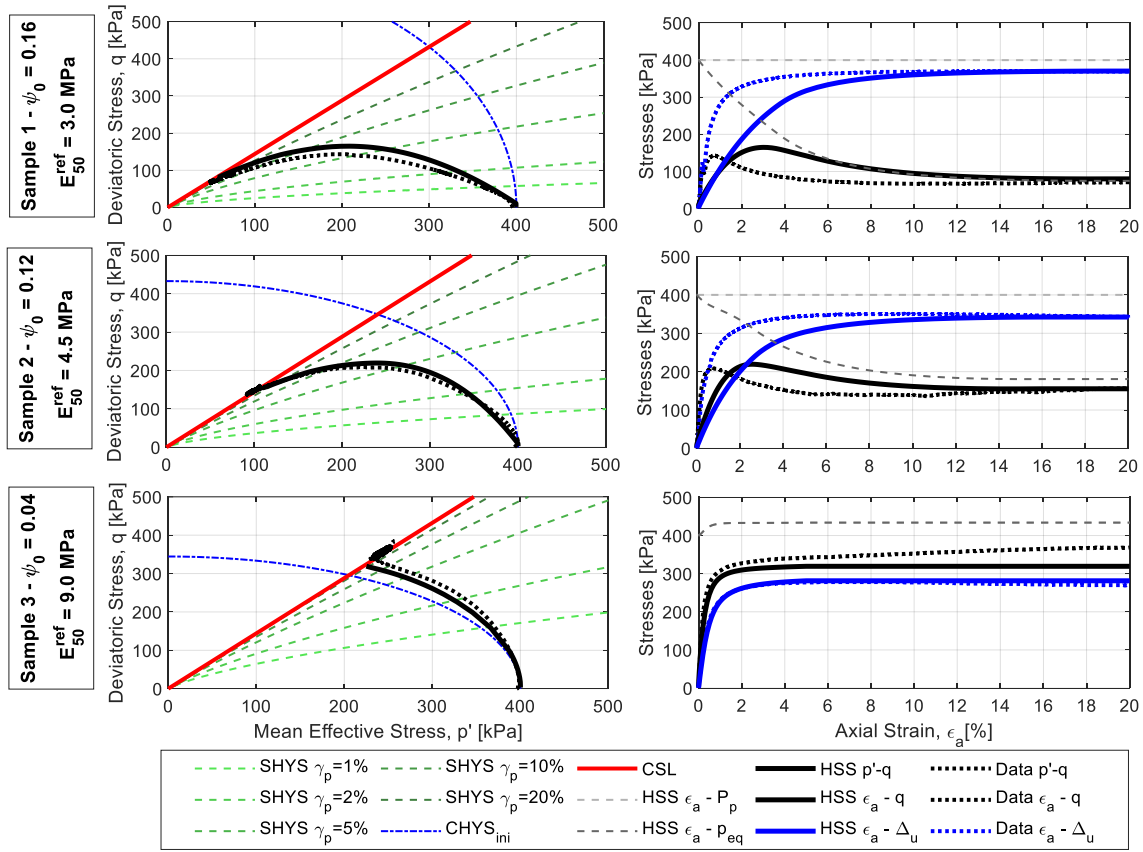

Figure 1. CIUC tests, comparison between data and simulations for samples 1, 2 and 3. SHYS is the yield surface in shear; CHYS is the cap yield surface.

### 2.4 Prediction for elemental CK0UC and MDSS tests

The comparison shown for CIUC tests validates the proposed calibration strategy, but isotropic pre-shearing stress states do not represent the expected in-situ conditions for an upstream-raised TSF. Moreover, the calibration is only shown for a unique pre-shearing mean effective stress of 400 kPa. To overcome this limitation, a parametric study is performed considering K0-consolidated stress before undrained shearing in triaxial and direct simple shear modes (i.e. CK0UC and DSS); in addition, two pre-shearing vertical effective stresses $\sigma'_{vo} = 200|400\ kPa$ are considered to evaluate if the response can be normalized, and so, if this calibration can be applied to all the in-situ stresses in the expected range.

The CK0UC and DSS elemental tests are performed using the HSS parameters shown in Table 1, with five different values for $E_{50}^{ref} = 3.0|4.5|6.0|7.5|9.0\ MPa$; note that three of them were calibrated from CIUC tests. A summary of the effective stress paths and stress-strain results is shown in Figure 2. It is observed that: i) the stress-strain response is qualitatively the same for the two initial effective stresses, anticipating that a normalized behavior is expected; ii) the residual shear strength reaches a steady value, typically for strains up to 30%; iii) the highest value of $E_{50}^{ref}$ (i.e. 9 MPa) entails no softening; as the parameter value is reduced, strain-softening becomes significant and both the peak and residual shear strengths decrease; iv) the strain at peak is lower than those reported in CIUC tests, which is expected as the initial K0 stress state is closer to failure.

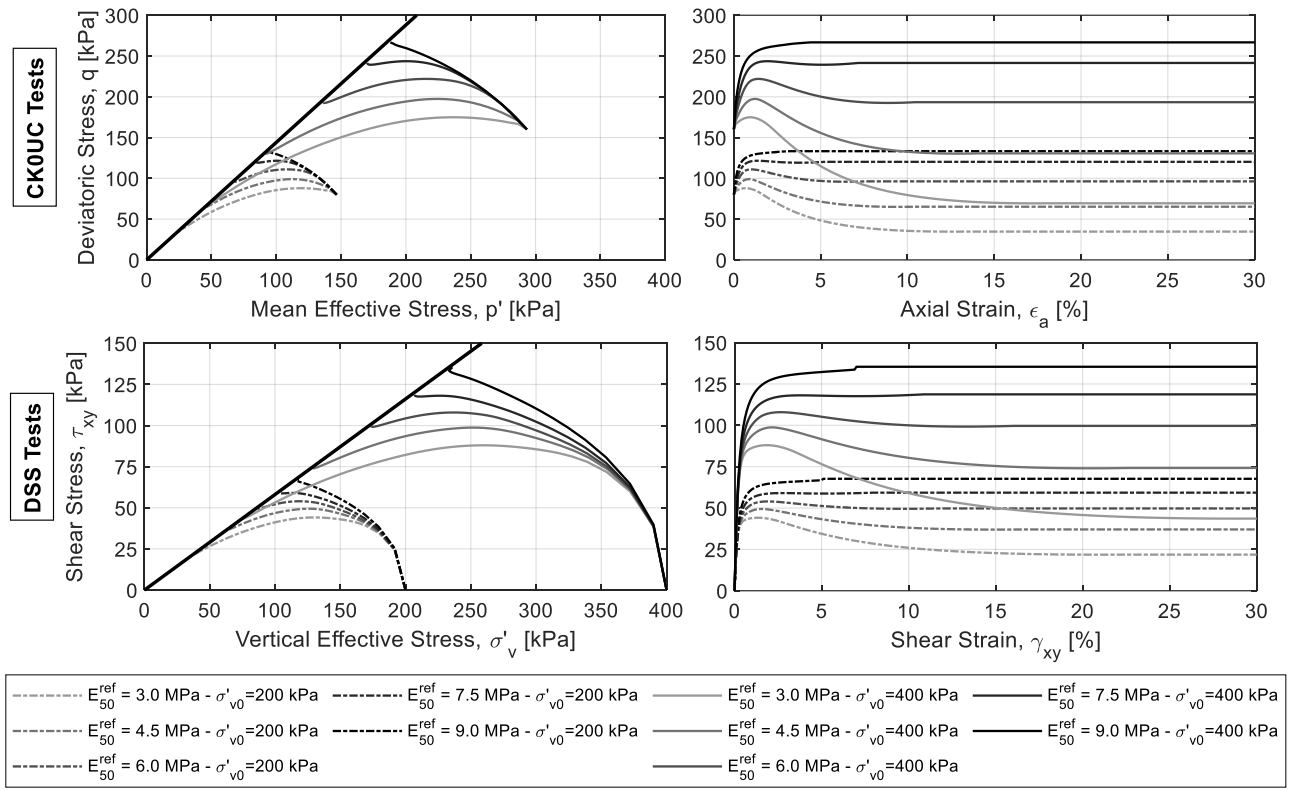

Figure 2. Parametric study of the effect of $E_{50}^{ref}$ on stress-strain response of CK0UC and DSS tests.

Peak and residual undrained shear strength ratios from CK0UC and MDSS simulations are calculated, as they are of paramount importance for static liquefaction assessment; results are summarized in Figure 3. It is shown that undrained shear strength ratios ($s_u/\sigma'_v$): i) have a normalized behavior; thus, they are independent on the initial effective stress; ii) are very similar for triaxial and direct simple shear modes; thus, HSS cannot capture strength anisotropy, which is expected as the model is isotropic; iii) can reach $s_u/\sigma'_v > 0.30$ for $E_{50}^{ref} = [7.5|9.0]\ MPa$, which might entail an strength overprediction for loose tailings; iv) entail peak value $s_u/\sigma'_v \cong 0.23$ and a residual value $s_{u,res}/\sigma'_v \cong 0.10$ in the worst scenario. It must be noted that the obtained peak and residual strength ratios for CK0UC and DSS tests are in agreement with the range reported by Jefferies & Been (2016); thus, in absence of high-quality laboratory data, the user might use CPTu data to estimate in-situ state parameter, chose a design value of peak|residual shear strength, and calibrate the HSS model to fit those values.

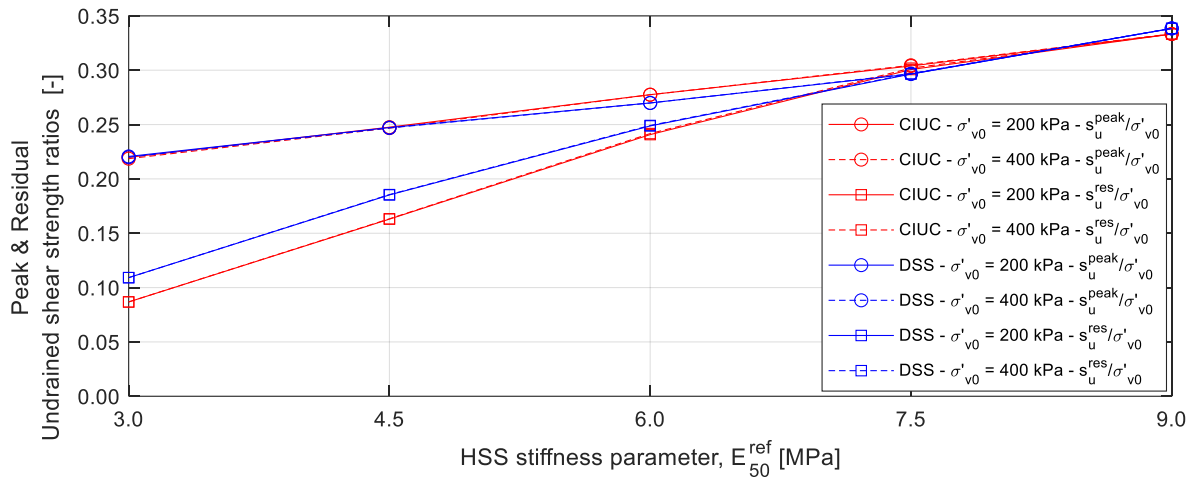

Figure 3. Peak and residual undrained shear strength ratios as a function of $E_{50}^{ref}$. Predictions from CK0UC and DSS elemental tests.

## 3 Static liquefaction numerical assessment of an upstream-raised TSF

### 3.1 Model description

An upstream-raised TSF is modelled in Plaxis 2D (2019) to evaluate its vulnerability against static liquefaction; the model geometry and mesh are presented in Figure 4. A horizontal contraction at the starter dam is chosen as a trigger mechanism; the aim is to represent hypothetical deformations due to: i) creep at the foundation, ii) excavation of the foundation material; iii) an accidental loss of material due to piping. The model has 4709 15-noded triangular elements and entails four geotechnical units: tailings, embankment raises, upper foundation and bedrock. Tailings are modelled using the parameters for the loosest sample. The embankment raises are modelled using HSS, the upper foundation is modelled as Mohr-Coulomb and the bedrock as linear elastic ($G$ is shear modulus in these models); parameters are shown in Table 2.

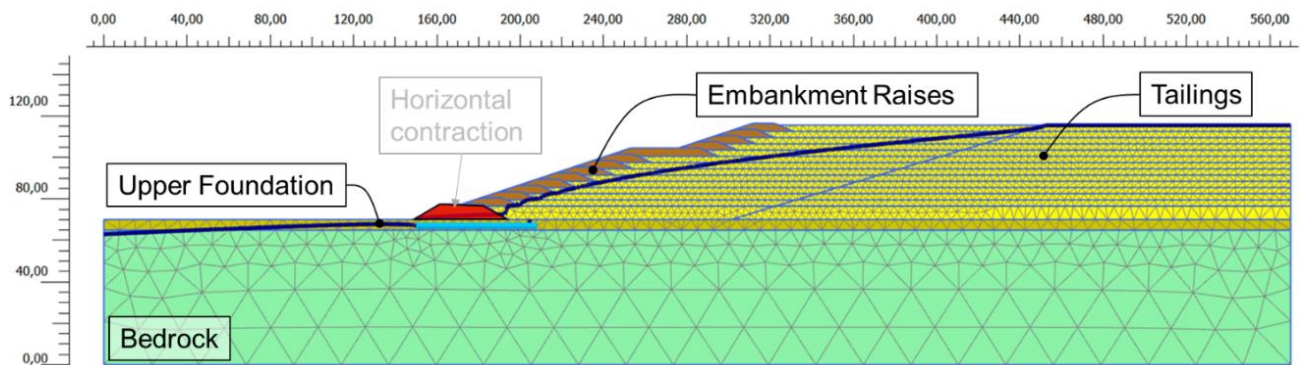

Figure 4. Plaxis 2D model. Geometry and mesh.

Table 2. HSS parameters for raises and foundation.

| Parameter | $c'$ | $\phi'$ | $G_0^{ref}|G$ | $\gamma_{0.7}$ | $p_{ref}$ | $R_f$ | $K_0^{NC}$ | $m$ | $\nu_{ur}$ | $E_{ur}^{ref}$ | $E_{oed}^{ref}$ | $E_{50}^{ref}$ |
|---|---|---|---|---|---|---|---|---|---|---|---|---|
| Units | kPa | ° | MPa | - | kPa | - | - | - | - | MPa | MPa | MPa |
| Tailings | 1 | 35 | 50 | 1E-4 | 100 | 0.875 | 0.60 | 0.55 | 0.25 | 60 | 9 | 3 |
| Emb. Raises | 5 | 33 | 205 | 1E-4 | 100 | 0.900 | 0.45 | 0.50 | 0.20 | 75 | 25 | 25 |
| Upper Found. | 1 | 40 | 4E5 | - | - | - | - | - | 0.20 | - | - | - |
| Bedrock | - | - | 4E5 | - | - | - | - | - | 0.20 | - | - | - |

### 3.2 Modelling strategy

The TSF is raised in several stages until reaching its current configuration, using an average raise height of 3.0 m and an average rate of rise of 2 m/year; the aim is to reasonable capture the staged construction process and its associated non-linearities that determine the in-situ stress field. A steady state groundwater flow is computed at each raise, while excess pore pressures are generated due to the tailings raise itself; thus, a decoupled flow-deformation scheme is adopted. Once the current TSF condition is reached, a horizontal contraction strain is gradually applied at the starter dam until reaching the global failure. It must be mentioned that this is conservatively done using a plastic-type calculation (i.e. no pore pressure dissipation at tailings) and ignoring suction (i.e. fully saturated above phreatic surface).

### 3.3 Results

Global failure is achieved for a horizontal contractive strain $\epsilon_{xx} = 1.3\%$ in the starter dam; in terms of horizontal displacements, this is approximately 6 cm towards downstream at the starter dam crest. Figure 5 shows the results; five gauss points are chosen to study the stress-strain-strength behavior along the failure. It is shown that: i) excess pore pressures are generated because of shear, as the higher magnitudes are concentrated along the shear band; ii) points B, C, D and E have similar stress ratios before the triggering event and show pre-peak hardening while strained, while point A has a higher initial stress ratio and shows no pre-peak hardening; iii) the effective stress paths are qualitatively in agreement to the analogous CK0UC results shown in Figure 2, for which the mobilized stress is very close to the undrained peak strength; iv) the undrained peak and residual shear strength ratios for points B, C, D and E are in agreement with those reported for the elemental tests ($s_u/\sigma'_v = 0.23$ and $s_{u,res}/\sigma'_v = 0.10$), which proves that the behavior captured by HSS can be normalized; however, point A shows higher peak and residual strength ratios, which can be attributed to its closeness to the foundation.

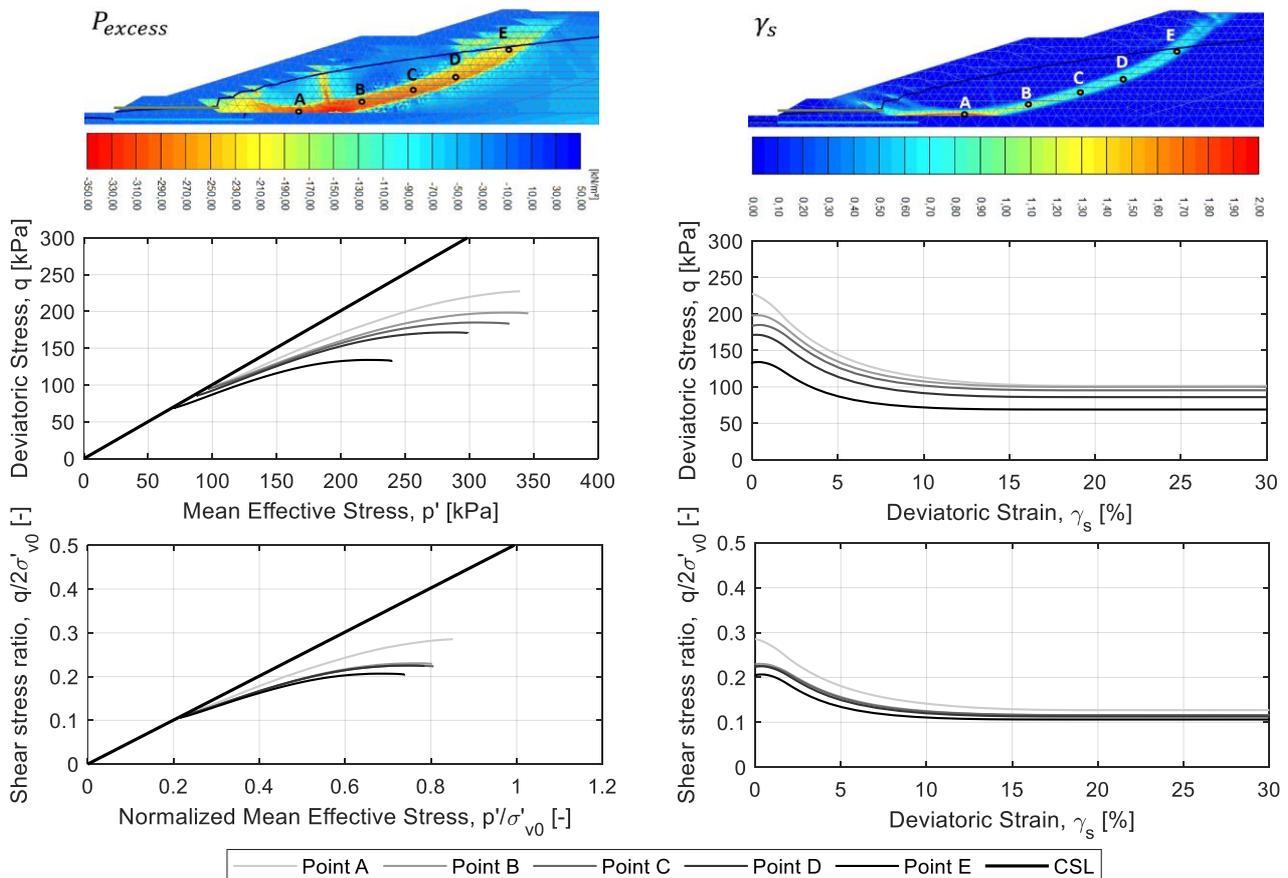

Figure 5. Upstream-raised TSF. Plaxis 2D results for a static liquefaction triggering analysis.

## 4  Conclusions

Existing methodologies to evaluate the stability of TSFs generally use limit equilibrium analyses considering peak and residual undrained shear strengths; these approaches neglect the work input required to drive the softening process that leads to progressive failure. This paper presents a simplified procedure to evaluate static liquefaction triggering of upstream-raised TSFs by means of finite element techniques and the well-known Hardening Soil model with small-strain stiffness (HSS).

A calibration methodology is proposed focusing on the stiffness parameters that control the rate of shear-induced plastic volumetric strains. The calibration process is shown for three CIUC tests conducted on reconstituted low plasticity silt-like tailings samples, with the same confining stress but different pre-shear void ratios; a set of parameters is calibrated to reasonable reproduce the undrained response of the material by just adjusting the parameter $E_{50}^{ref}$. Results are validated by a parametric study of elemental CK0UC and DSS tests, proving that the response can be normalized by the pre-shear effective stress state and that the achieved peak and undrained shear strength ratios agree with the values reported in the literature. Thus, in absence of high-quality laboratory data, the user might use CPTu data to estimate in-situ state parameters, chose a design value of peak | residual shear strength, and calibrate the HSS model to fit those values.

Finally, an application to a real TSF is presented to evaluate its vulnerability to liquefy for a trigger mechanism of lateral spreading due to a displacement of the starter dam. The staged construction of the facility is modelled in detail to capture the associated non-linearities that determine the in-current situ stress field; a horizontal contraction strain is gradually applied at the starter dam until reaching the global failure due to static liquefaction. In the example, results show that failure is triggered after a horizontal displacement of 6 cm towards downstream at the starter dam crest. The stress-strain response of five gauss points located along the failure is studied: most of them show a slight pre-peak hardening and a normalizable behavior with peak and residual undrained shear strength ratios that agree well with the calibration exercise.